\documentclass[pra,twocolumn,superscriptaddress,showpacs,floatfix,longbibliography]{revtex4-2}
\usepackage{mathrsfs,braket}
\usepackage{amssymb, amsbsy, amsmath, latexsym, dsfont, array, layout,
graphicx,mathrsfs,color,ulem,bm}
\usepackage[colorlinks=true,citecolor=blue,urlcolor=blue]{hyperref}
\usepackage{siunitx}

\begin{document}

\title{Chiral state transfer under dephasing}

\author{Konghao Sun}
\affiliation{CAS Key Laboratory of Quantum Information, University of Science and Technology of China, Hefei 230026, China}
\author{Wei Yi}
\email{wyiz@ustc.edu.cn}
\affiliation{CAS Key Laboratory of Quantum Information, University of Science and Technology of China, Hefei 230026, China}
\affiliation{CAS Center For Excellence in Quantum Information and Quantum Physics, Hefei 230026, China}

\begin{abstract}
Exceptional points emerge in the complex eigenspecra of non-Hermitian systems, and give rise to rich critical behaviors. An outstanding example is the chiral state transfer, where states can swap under an adiabatic encircling around the exceptional point, but only along one direction.
In dissipative quantum systems, such exceptional-point encirclings are often accompanied by decoherence, whose impact is
beyond the description of non-Hermitian Hamiltonians.
In this work, we study in detail the effects of dephasing on the encircling dynamics, adopting the full Lindblad master equation.
Introducing experimentally relevant quantum-jump processes that account for dephasing, we show that gaps emerge in the eigenspectra landscape of the corresponding Liouvillian superoperator. It follows that the chiral state transfer does not take place in the adiabatic limit, since the system always adiabatically follows the quasi-steady state of the Liouvillian regardless of the encircling direction.
Nevertheless, the chirality is restored at intermediate encircling times, where the dynamics is non-adiabatic in both encircling directions, distinct from the typical chiral state transfer in non-Hermitian systems.
While our results are applicable to several recent experiments, we examine a recent cold-atom experiment in particular, and show that the observed long-time chirality is but limited to the special encircling path therein.
Our study provides further insight into the chiral state transfer under experimental conditions, and is helpful for controlling open-system dynamics from the perspective of non-Hermitian physics.
\end{abstract}

\maketitle

\section{Introduction}

An open quantum system exchanges energy, particle, or information with its environment, thus defies the description of Hermitian Hamiltonians.
A convenient, if phenomenological, alternative is provided by the non-Hermitian framework~\cite{ref1}.
With complex eigenspectra and ramified symmetries~\cite{ref2}, non-Hermitian Hamiltonians exhibit rich and exotic features, including the parity-time symmetry~\cite{pt1,pt2,pt3,pt4,pt5,pt6,pt7,pt8,pt9,pt10,pt11,pt12}, the non-Hermitian nodal phases \cite{nod1,nod2,nod3,nod4,nod5}, and non-Hermitian topology~\cite{topo1,topo2,topo3,topo4,topo5,topo6,topo7,topo8,topo9,topo10,topo11,topo13,topo14,topo15,topo16}.
These phenomena have stimulated intense research interest in recent years, not only because of
the fresh insights they offer regarding open systems, but also for their potential applications.

The recent surge of interest in non-Hermitian physics can be traced back to the discovery of the parity-time symmetry~\cite{pt1,pt2}, under which a non-Hermitian Hamiltonian can acquire completely real eigenspectrum. At the critical point where the symmetry becomes spontaneously broken, eigenenergies and eigenstates coalesce, as the eigenspectrum transits from being completely real to complex.
These so-called exceptional points (EPs), while absent in Hermitian systems, generally exist in non-Hermitian settings, even without the parity-time symmetry~\cite{realspec}. Their sensitive dependence on the tuning parameters and spectral landscapes are promising for enhanced sensing~\cite{sen1,sen2,sen3} and quantum-device design~\cite{dev1,dev2,dev3}.
In the latter case, a particularly useful feature is the chiral state transfer near an EP~\cite{ec1,ec2,ec3,ec4,ec5,ec6}.
By slowly varying the parameters on a closed loop near the EP (encircling the EP preferred but not necessary), the dynamics could lead to a state exchange, depending on the direction of the encircling---the state is switched only one way around.
Though seemingly contradictory to the adiabatic theorem, the chiral state transfer is a consequence of the complex eigenspectral structure in the parameter space, together with the path-dependent amplification of the non-adiabatic couplings~\cite{ec5}.

Experimentally, the chiral state transfer has been observed in classical systems with gain and loss~\cite{cla1,cla2}, and in dissipative quantum settings of photons~\cite{phot1}, superconducting qubits~\cite{sq1,sq2}, trapped ions~\cite{ions}, solid spins~\cite{nv}, and cold atoms~\cite{ca}.
In quantum systems, the non-Hermiticity is realized by enforcing postselection~\cite{ps1,ps2}. More specifically, most of these open quantum systems feature weak interactions with a Markovian environment, with the full, unconditional dynamics described by a Lindblad master equation.
The time evolution of the Lindblad master equation can be understood as the average of infinitely many quantum trajectories, each represented by a stochastic wave function, whose time evolution is driven by a non-Hermitian effective Hamiltonian and interrupted
by quantum jumps. Hence, non-Hermiticity naturally emerges for trajectories where quantum jumps are absent.
Enforcing postselection constitutes picking out these trajectories, leading to the conditional dynamics governed by the non-Hermitian effective Hamiltonian alone.
However, realistic open systems often consist of many degrees of freedom, wherein additional quantum-jump processes are inevitable.
In a series of recent experiments~\cite{sq1,sq2,ca}, for instance, it has been demonstrated that the EP-induced chiral state transfer persists despite these additional quantum jumps.
This is perhaps not too surprising, if the quantum-jump processes are to be glossed as perturbations whose impact on the non-Hermitian dynamics manifests only as decoherence or heating.
But from the perspective of open quantum systems, it is more appropriate to analyze the dynamics using the full Lindblad master equation, where the time evolution of the density matrix is driven by the Liouvillian superoperator. Since the Liouvillian is intrinsically non-Hermitian and can host its own EPs~\cite{sq1,sq2,LEP}, it is natural to pose the question whether the impact of the quantum jumps on the EP-induced chiral state exchange can be understood from the Liouvillian spectra.
The question is non-trivial, not least because the spectral landscape of Liouvillians, along with the Liouvillian EPs, are in general distinct from those of the corresponding non-Hermitian Hamiltonians.
Dynamics can therefore be quite different when the decoherence-inducing quantum jumps are present.

In this work, we focus on the chiral state transfer under quantum-jump processes that lead to dephasing. These terms are relevant in recent experiments of EP encirclings in superconducting qubits~\cite{sq1,sq2} and cold atoms~\cite{ca}. We start from the full Lindblad master equation, and demonstrate that
the presence of quantum-jump processes open up a finite gap in the eigenspectra landscape of the Liouvillian. A direct consequence is that the chiral state transfer does not occur when the encircling time approaches infinity. When the encircling time is long enough, the system always ends up in a quasi-steady state of the Liouvillian, regardless of the encircling direction. While the chirality of the state transfer can be restored at intermediate encircling times,
the dynamics then is distinct from that under the non-Hermitian Hamiltonian.
In the non-Hermitian case, the encircling dynamics is adiabatic in one direction and non-adiabatic in the other, and the final state is a pure state.
By contrast, under dephasing and at intermediate times, the dynamics in either direction is non-adiabatic, and the final state is mixed.

In light of these understandings, we discuss the recent EP-encircling experiment in cold atoms~\cite{ca}. Therein, non-Hermiticity is implemented by postselecting the internal states of atoms undergoing laser-induced spontaneous decay. However, branching of the spontaneous decay necessarily gives rise to additional quantum-jump processes, causing dephasing and heating. While the experiment reports chiral state transfer under a fairly long encircling time, we show that this is facilitated by the particular choice of the encircling path in the experiment. Therein, the path consists of a segment where the dynamics is driven by a purely Hermitian Hamiltonian. The Liouvillian gap closes along this sector, such that the dynamics is decidedly non-adiabatic on the Liouvillain eigenspectral landscape in either encircling direction.
We demonstrate that, by choosing a more general encircling path, the chiral state transfer occurs only at intermediate encircling times, consistent with our prediction above.
The effect of the Liouvillian-gap closing along the Hermitian segment can therefore be understood as pushing the chirality condition from the intermediate time to the infinite-time limit.

Our work is organized as follows. In Sec.~II, we present the four-level system that we consider, inspired by the recent cold-atom experiment. We analyze the encircling dynamics in detail in Sec.~III. In Sec.~IV, we discuss the experiment~\cite{ca} from the perspective of the Liouvillian eigenspectrum, and consider a general encircling path where the results in Sec.~III can be recovered. We conclude in Sec.~V.

\begin{figure}[tbp]
	\includegraphics[scale=1]{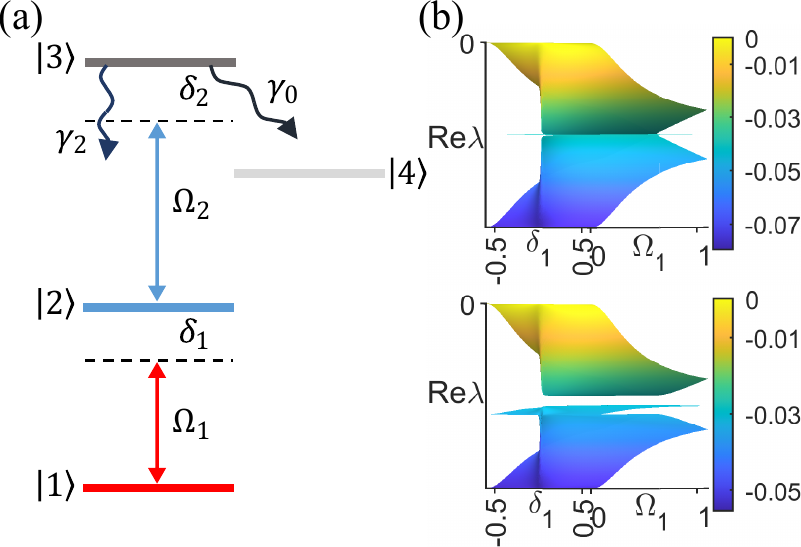}%
	\caption{(a) Schematic illustration of the four-level system considered in this work.
States $|1\rangle$ and $|2\rangle$ are coupled with a coupling rate $\Omega_1$ and detuning $\delta_1$. Dissipation is introduced by coupling states $|2\rangle$ and $|3\rangle$ with a rate $\Omega_2$ and detuning $\delta_2$. The spontaneous decay rates for $|3\rangle\rightarrow |4\rangle$ and $|3\rangle\rightarrow |2\rangle$ are denoted, respectively, as $\gamma_0$ and $\gamma_2$.
(b) The Liouvillian spectra $\text{Re}\lambda$ without (upper panel) and with (lower pannel) the recycling term $L_{\phi}\rho L^\dag_{\phi}$, respectively (see main text for the definition of $L_\phi$). A Liouvillian gap is visible in the lower panel. The parameters in (b): $\gamma_0=50$, $\Omega_2=1$, and $\delta_2=0$. We take $\gamma_2=0$ ($\gamma_2=10$) for upper (lower) panel.} \label{fig1}
\end{figure}

\begin{figure*}[tbp]
	\includegraphics[scale=1]{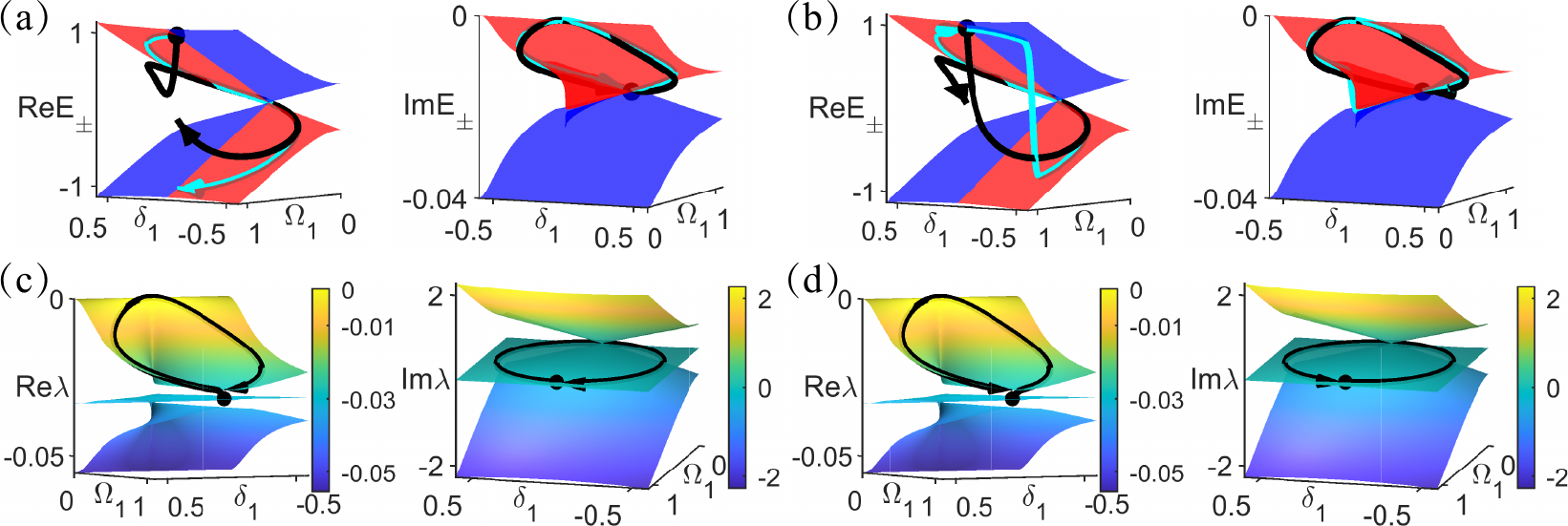}%
	\caption{Trajectories of the encircling dynamics in the adiabatic limit, for (a)(c) clockwise rotations and (b)(d) counterclockwise rotations.
(a)(b) Trajectories (black and cyan) against the eigenspectra of the non-Hermitian Hamiltonian $H_{\text{eff}}$. The red (blue) branch indicates the eigenstate with a larger (smaller) imaginary component, and hence smaller (larger) loss.
The black trajectory corresponds to the dynamics under the master equation, calculated using Eq.~(\ref{eq:mastertraj}).
The cyan trajectory corresponds to the ideal EP-encirlcing dynamics under $H_{\text{eff}}$ alone, calculated using Eq.~(\ref{eq:nonHtraj}).
(c)(d) Trajectories against the eigenspectra of the Liouvillian $\mathcal{L}$. The black trajectory is calculated according to Eq.~(\ref{eq:mastertraj2}).
For all cases, the initial state is an eigenstate of $H_{\text{eff}}$, and the encircling path is $\delta_1\left(t\right)=0.5\sin{\left(\pm 2\pi t/T\right)}$ and $\Omega_1\left(t\right)=0.5+0.5\cos{\left(\pm 2\pi t/T\right)}$, with the total encircling time $T=15000$. Other parameters are $\Omega_2=1$, $\gamma_0=50$, $\gamma_2=10$, and $\delta_2=0$.} \label{fig2}
\end{figure*}

\begin{figure*}[tbp]
	\includegraphics[scale=1]{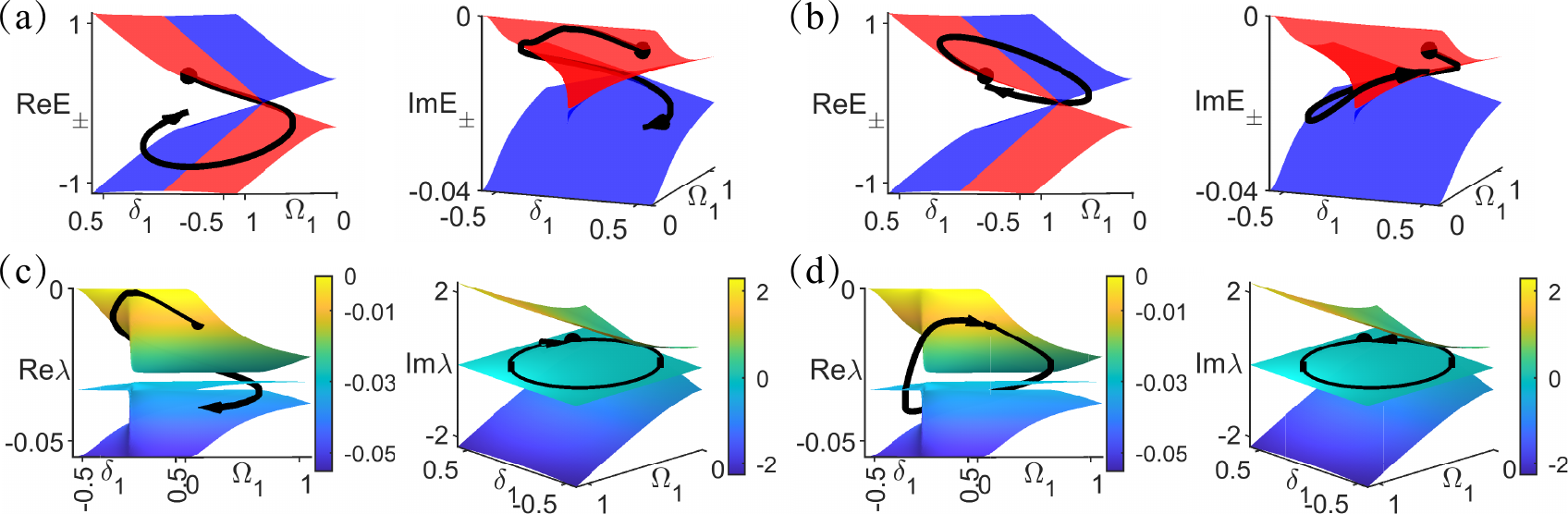}%
	\caption{Trajectories of the encircling dynamics for an intermediate encircling time $T=150$, for (a)(c) clockwise rotations and (b)(d) counterclockwise rotations.
(a)(b) Trajectories against the eigenspectra of the non-Hermitian Hamiltonian $H_{\text{eff}}$, calculated according to Eq.~(\ref{eq:mastertraj}).
(c)(d) Trajectories against the eigenspectra of the Liouvillian $\mathcal{L}$, calculated according to Eq.~(\ref{eq:mastertraj2}).
The encircling path is $\delta_1\left(t\right)=0.5\sin{\left(\pm 2\pi t/T+2\pi/3\right)}$ and $\Omega_1\left(t\right)=0.5+0.5\cos{\left(\pm 2\pi t/T+2\pi/3\right)}$. Other parameters are the same as those in Fig.~\ref{fig2}.
\label{fig3}}
\end{figure*}

\section{Model}
Motivated by the recent experiments~\cite{sq1,sq2,ca}, we consider a four-level system illustrated in Fig.~\ref{fig1}. The two-level open system is encoded in the states $\{|1\rangle,|2\rangle\}$, which are coupled to each other. For simplicity, we assume the coupling does not involve momentum transfer, so that the external degrees of freedom are decoupled. A laser-induced loss is imposed on state $|2\rangle$ by coupling it to an excited state $|3\rangle$, which undergoes spontaneous decay. We consider the case where the decay has two different channels, one into a bystander state $|4\rangle$, the other back into state $|2\rangle$. Adiabatically eliminating state $|3\rangle$ and projecting out state $|4\rangle$, we derive the Lindblad master equation governing the dynamics of the open system (see Appendix for a detailed derivation where we take $\hbar=1$)
\begin{align}
	\dot{\rho}=-i(H_{\text{eff}}\rho-\rho H_{\text{eff}}^{\dagger})+L_{\phi}\rho L_{\phi}^{\dagger}-\frac{1}{2}L_{\phi}^\dag L_{\phi}\rho-\frac{1}{2}\rho L_{\phi}^\dag L_{\phi}, \label{eq1}
\end{align}
where the non-Hermitian effective Hamiltonian is
\begin{align}
	H_{\text{eff}}=H_0-i\Gamma |2\rangle\langle 2|.\label{eq:nonH}
\end{align}
In the basis of $\{|1\rangle,|2\rangle\}$, the matrix for the Hermitian part of the Hamiltonian is
\begin{align}
H_0=\begin{bmatrix}
	\begin{array}{cc}
			\frac{\delta_1}{2} & -\Omega _1 \\
			-\Omega _1^* & -\frac{\delta_1}{2}
		\end{array}
\end{bmatrix},
\end{align}
where the coupling rate $\Omega_1$ and the detuning $\delta_1$ are illustrated in Fig.~\ref{fig1}.
The quantum-jump operator $L_\phi=\sqrt{\gamma_{\phi}}\left|2\right\rangle\left\langle 2\right|$ accounts for the spontaneous decay
$|3\rangle\rightarrow |2\rangle$, which leads to the dephasing of the $\{|1\rangle,|2\rangle\}$ subsystem.
The non-Hermitian term in Eq.~(\ref{eq:nonH}) characterized by $\Gamma$ originates from the decay $|3\rangle\rightarrow|4\rangle$, and is enforced through postselection in the experiment by considering dynamics of atoms in the states $|1\rangle$ and $|2\rangle$ only~\cite{ca,luoleexp}.
Explicit expressions for $\gamma_{\phi}$ and $\Gamma$ in terms of the control paramters $\Omega_{1,2}$, $\delta_{1,2}$, and $\gamma_{0,2}$ (all illustrated in Fig.~\ref{fig1}) are summarized in the Appendix.

In the absence of the decay process $|3\rangle\rightarrow |2\rangle$, the dynamics of the subsystem $\{|1\rangle,|2\rangle\}$ is driven by the non-Hermitian effective Hamiltonian $H_{\rm{eff}}$. It is based on similar non-Hermitian Hamiltonians that EP encircling and chiral state transfer were studied in recent experiments~\cite{ca}. The decay channel $|3\rangle\rightarrow |2\rangle$, besides modifying $\Gamma$ and introducing an additional non-Hermitian contribution $\frac{1}{2}L_{\phi}^\dag L_{\phi}$ to $H_{\rm{eff}}$, gives rises to the quantum recycling term $L_{\phi}\rho L^\dag_{\phi}$ that is responsible for dephasing (see Appendix).

The impact of dephasing is further visible on the eigenspectral landscape of the Liouvillian superoperator $\mathcal{L}$. Specifically, vectorizing the density matrix
\begin{align}
	\rho=\begin{bmatrix}
	    \begin{array}{cc}
	    	\rho_{11} & \rho_{12}\\
	    	\rho_{21} & \rho_{22}
	    \end{array}
	\end{bmatrix}
    \rightarrow
    \begin{pmatrix}
    	\rho_{11}\\
    	\rho_{12}\\
    	\rho_{21}\\
    	\rho_{22}
    \end{pmatrix},
\end{align}
we cast the Liouvillian in the matrix form
\begin{align}
	\mathcal{L}
	=
	\begin{bmatrix}
		\begin{array}{cccc}
			0 & -i\Omega_1^{*}  & i\Omega _1 & 0 \\
			-i\Omega_1 & -i\delta_1-\frac{\left|\Omega_2\right|^2 \gamma^3}{\Delta^4} & 0 & i\Omega_1 \\
			i\Omega_1^{*} & 0 & i\delta_1-\frac{\left|\Omega_2\right|^2 \gamma^3}{\Delta^4} & -i\Omega_1^{*}\\
			0 & i\Omega_1^{*} & -i\Omega_1 & -\frac{\left|\Omega_2\right|^2 \gamma_0 \gamma^2}{\Delta^4}\\
		\end{array}
	\end{bmatrix}.
\end{align}
The density-matrix dynamics is then captured by $\dot{\rho}=\mathcal{L}\rho$. The eigenspetrum $\lambda$ of $\mathcal{L}$ is defined as $\mathcal{L}\rho=\lambda\rho$.
In Fig.~\ref{fig1}(b), we show the real component of the Liouvillian spectrum $\text{Re}\lambda$ in the parameter space of $(\Omega_1,\delta_1)$.
The dynamics is not trace-preserving, because of the population of state $|4\rangle$ which is outside of the two-level open system. This is manifested in the Liouvillian spectra as the deviation of the high-lying sheet from $\text{Re}\lambda=0$. Since $\text{Re}\lambda=0$ corresponds to the steady state of tracing-preserving open-system dynamics, we regard the high-lying sheet in Fig.~\ref{fig1}(b) as the quasi-steady state.
Importantly, a Liouvillian gap emerges as soon as the recycling term is switched on. As illustrated in the lower panel of Fig.~\ref{fig1}(b), the gap separates the quasi-steady state from other branches, thus impacting the long-time dynamics of the open system.

\begin{figure}[tbp]
	\includegraphics[scale=1]{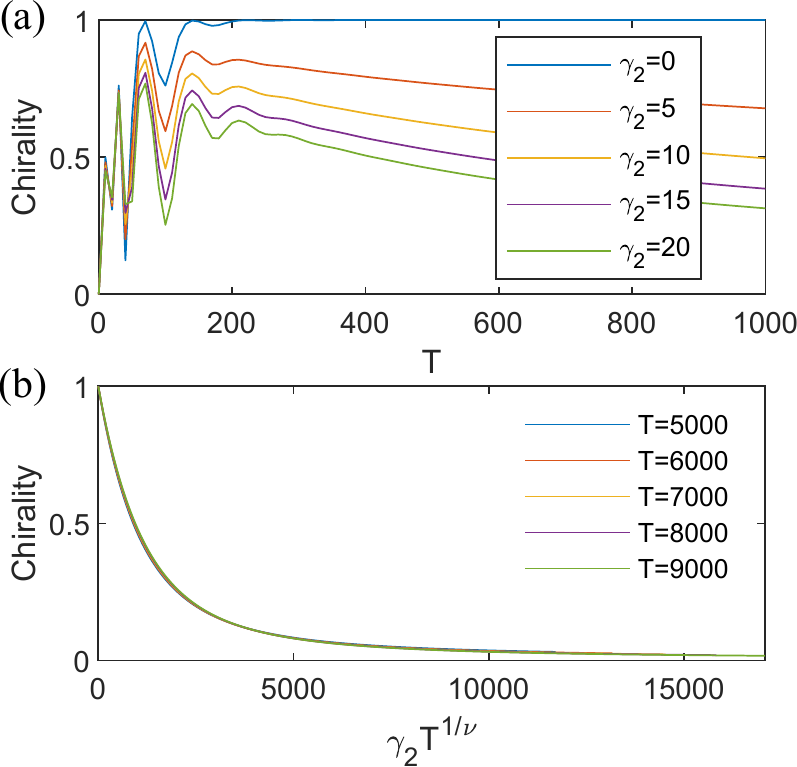}%
	\caption{(a) Chirality as a function of the total encircling time $T$ under different values of $\gamma_2$. (b) Scaling relation of the chirality with respect to $\gamma_2$ and $T$. The initial state is $\rho\left(0\right)=\left|1\right\rangle\left\langle1\right|$. We determine $\nu\approx1.561$ from numerical fitting (see main text).}
\label{fig4}
\end{figure}

\section{EP encircling under dephasing}

In the absence of dephasing ($\gamma_2=0$), an EP exists at $\{\delta_1=0, \Gamma=2|\Omega_1|\}$ under $H_{\text{eff}}$ alone. When the system parameters are adiabatically tuned in a closed loop near the EP, depending on the encircling direction, the time-evolved state can be switched or remain unchanged, on returning to the initial parameters. This is illustrated in Fig.~\ref{fig2}(a)(b) by the cyan trajectories, calculated according to
\begin{align}
	\bar{E}_0(t)=\frac{\sum_{i=\pm}|\langle\chi_i^L\left(t\right)|\psi_0\left(t\right)\rangle|^2 E_i\left(t\right)}{\sum_{i=\pm}|\langle\chi_i^L\left(t\right)|\psi_0\left(t\right)\rangle|^2},\label{eq:nonHtraj}
\end{align}
where $\langle\chi_i^L|$ ($i=\pm$) are left eigenstates of $H_{\text{eff}}$, with $H^\dag_{\text{eff}}|\chi_i^L\rangle=E^\ast_i |\chi_i^L\rangle$ and $E_{\pm}$ are the eigenvalues of $H_{\rm{eff}}$.
The time-evolved state $|\psi_0\left(t\right)\rangle$ is given by
\begin{align}
	i\frac{d}{d t}|\psi_0\left(t\right)\rangle=H_\text{eff}\left(t\right)|\psi_0\left(t\right)\rangle.
\end{align}
Such a chiral state transfer originates from the interplay of the topology of the eigenspectra landscape of the Hamiltonian in the parameter space, and the non-Hermitian amplification of the non-adiabatic coupling terms. As analyzed in previous theoretical studies and confirmed in various experiments, the dynamics adiabatically follows the eigenspectral landscape (the state gets switched) in one direction, and is non-adiabatic in the other (the state remains the same).
Here the chirality is signaled by the different final states for different encircling directions.
However, this is no longer the case when $\gamma_2\neq 0$.

In the upper panels of Fig.~\ref{fig2}(a)(b), we show, respectively, the trajectories (in black) for the clockwise and counterclockwise encirclings in the presence of dephasing.
We evolve the density matrix according to Eq.~(\ref{eq1}), and calculate the trajectories on the eigenspectra landscape $E_{\pm}$. Note that, throughout our work, we calculate the eigenspectra of $H_{\rm{eff}}$ by setting $\gamma_2=0$.
Specifically, the trajectories in black are calculated according to
\begin{align}
\bar{E}(t)=\frac{\sum_{i=\pm}\langle\chi_i^L\left(t\right)|\rho\left(t\right)|\chi_i^L\left(t\right)\rangle E_i\left(t\right)}{\sum_{i=\pm}\langle\chi_i^L\left(t\right)|\rho\left(t\right)|\chi_i^L\left(t\right)\rangle},\label{eq:mastertraj}
\end{align}
The key observation from Fig.~\ref{fig2}(a)(b) is that, when the recycling term (or dephasing) is switched on, the chirality disappears---the state always returns to
the same final state in the long-time limit, regardless of the encircling direction.

Such a behavior can be straightforwardly understood from the trajectories on the eigenspectra landscape of the Liouvillian. This is shown in Fig.~\ref{fig2}(c)(d), where the trajectories (in black) are calculated according to
\begin{align}
\bar{\lambda}(t)=\frac{\sum_{i=1}^4|\langle\phi_i^L\left(t\right)|\psi\left(t\right)\rangle|^2 \lambda_i\left(t\right)}{\sum_{i=1}^4|\langle\phi_i^L\left(t\right)|\psi\left(t\right)\rangle|^2},\label{eq:mastertraj2}
\end{align}
where $|\psi\left(t\right)\rangle$ is the vectorized density matrix $\rho\left(t\right)$, and $|\phi_i^L\left(t\right)\rangle$ satisfies $\mathcal{L}^\dag\left(t\right)|\phi_i^L\left(t\right)\rangle=\lambda_i^\ast |\phi_i^L\left(t\right)\rangle$. Apparently, because of the presence of the dephasing-induced Liouvillian gap, the state first relaxes to and then adiabatically follows the quasi-steady state of the Liouvillian (high-lying sheet). The chirality thus vanishes.

\begin{figure*}[tbp]
	\includegraphics[scale=1]{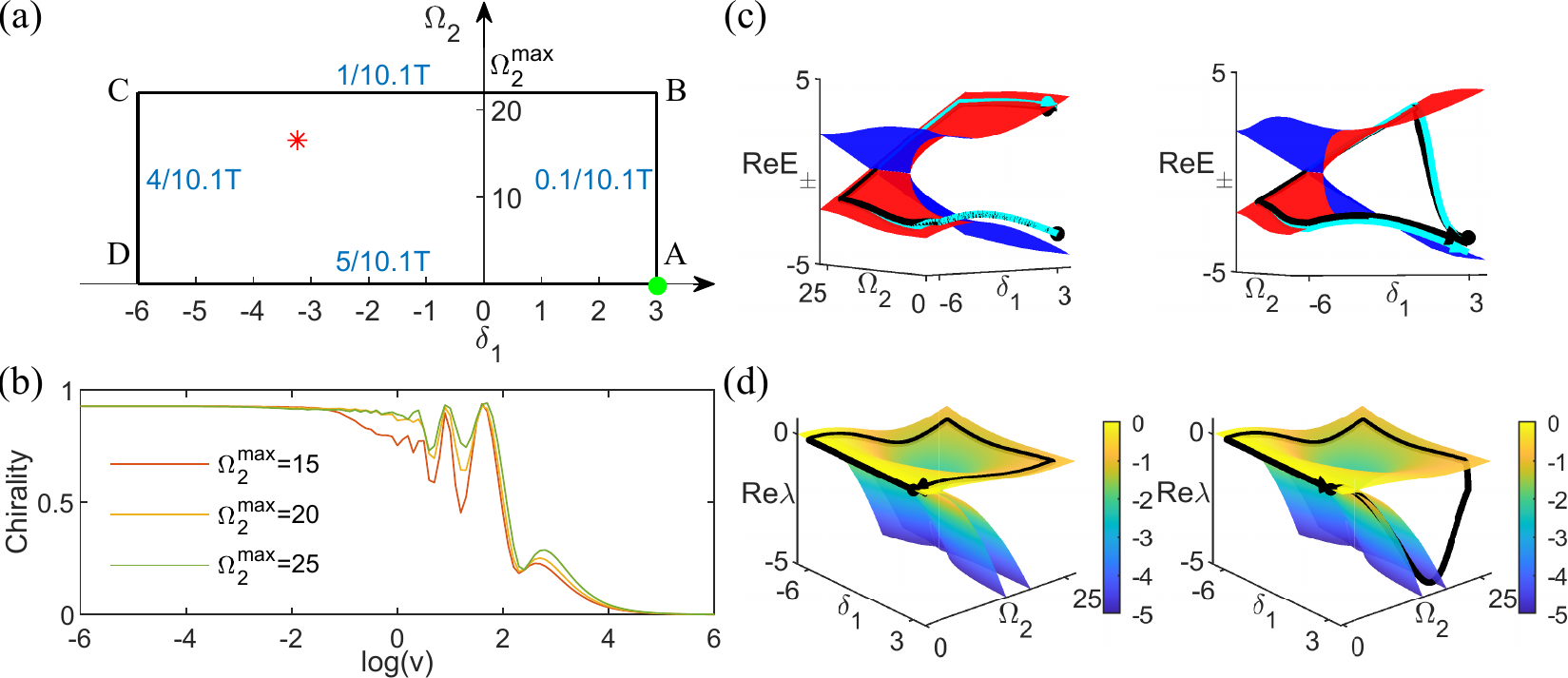}%
		\caption{(a) Encircling path adopted in the experiment~\cite{ca}. For the various symbols, we adopt the convention in Ref.~\cite{ca}: red star is the EP of $H_{\text{eff}}$ in the parameter space, the green dot is the initial point of encircling, and the blue numbers are the ratio of time spent on each segment. (b) Chirality as a function of $\log(v)$ for various $\Omega_2^{\text{max}}$. Here $\Omega_2^{\text{max}}$ is the maximum $\Omega_2$ on the encircling path. (c)(d) Trajectories against the real eigenspectra of the non-Hermitian Hamiltonian $H_{\text{eff}}$ (c) and the Liouvillian $\mathcal{L}$ (d) for clockwise rotations (left panel) and counterclockwise rotations (right panel) for $\log(v)=-2.66$. In (c), the black and cyan trajectories are calculated according to Eq.~(\ref{eq:mastertraj}) and Eq.~(\ref{eq:nonHtraj}), respectively. In (d), the black trajectory is calculated according to Eq.~(\ref{eq:mastertraj2}).
 \label{fig5}}
\end{figure*}

However, when the total encircling time $T$ is finite, the chiral behavior can emerge even under dephasing, though the dynamics is always non-adiabatic regardless of the encircling direction. This is illustrated in Fig.~\ref{fig3}, where we choose an intermediate encircling time. Notice how the trajectories always involve non-adiabatic jumps, whether projected onto the spectral landscape of $H_{\rm{eff}}$ [as in Fig.~\ref{fig3}(a)(b)], or that of the Liouvillian $\mathcal{L}$ [as in Fig.~\ref{fig3}(c)(d)].
This scenario is actually what has been observed in Refs.~\cite{sq1,sq2}, as
we have numerically checked that the reported chiral state transfer would disappear at longer encircling times.

To quantify our observations above, we adopt the definition of chirality~\cite{sq2}
\begin{align}
C=\frac{1}{2}\text{Tr}\left[\sqrt{\left(\widetilde{\rho}_{\text{cw}}-\widetilde{\rho}_{\text{ccw}}\right)^{\dagger}\left(\widetilde{\rho}_{\text{cw}}-\widetilde{\rho}_{\text{ccw}}\right)}\right].
\end{align}
Here $\widetilde{\rho}_{\text{cw},\text{ccw}}=\rho_{\text{cw},\text{ccw}}/\rm{Tr}\left(\rho_{\text{cw},\text{ccw}}\right)$, with $\rho_{\text{cw},\text{ccw}}$ being the final-time density matrix of the clockwise (cw) and counterclockwise (ccw) encircling.
The chirality $C$ takes values in between $0$ and $1$, and represents the distinguishability between the final density matrices for clockwise and counterclockwise encirclings. We have $C=0$ when the two density matrices are the same, and $C=1$ when they are orthogonal pure states.

We show the calculated chirality as a function of the encircling time $T$, for different $\gamma_2$ in Fig.~\ref{fig4}(a). In the absence of dephasing ($\gamma_2=0$), $C$ approaches a finite value close to unity in the long-time limit. By contrast, as long as $\gamma_2$ is finite, $C$ always approaches zero for a sufficiently long encircling time.
Another prominent feature of Fig.~\ref{fig4}(a) is the emergence of a chirality peak at intermediate $T$, which corresponds to the parameter regime of the chiral transfer illustrated in Fig.~\ref{fig3}.
Further, the final-time chirality manifests a universal scaling with $\gamma_2$ and $T$, as shown in Fig.~\ref{fig4}(b), with the relation $C=f(\gamma_2 T^{1/\nu})$ where $\nu\approx1.561$ and $f(x)$ is a universal scaling function.

Hence, along a general encircling path, the chiral state transfer only emerges at intermediate encircling times under dephasing, which is qualitatively different from the EP-encircling dynamics in a purely non-Hermitian system.
In the long-time limit, dephasing opens up a Liouvillian gap, the dynamics is essentially adiabatic on the Liouvillian spectral landscape, regardless of the encircling direction.
The conclusions above should apply to a series of recent experiments~\cite{sq1,sq2,ca}, where EP encircling has been studied in the presence of dephasing. Using numerical simulations, we have checked that the chiral behavior reported in Refs.~\cite{sq1,sq2} indeed occurs only at intermediate encirlcing times, and disappears in the long-time limit.
Nevertheless, such a conclusion seems to contradict the observations of a recent cold-atom experiment~\cite{ca}, where the chiral state transfer is observed at sufficiently long encircling time. In the following, we show that such an observation is based on the special encircling path adopted therein.

\begin{figure*}[tbp]
	\includegraphics[scale=1]{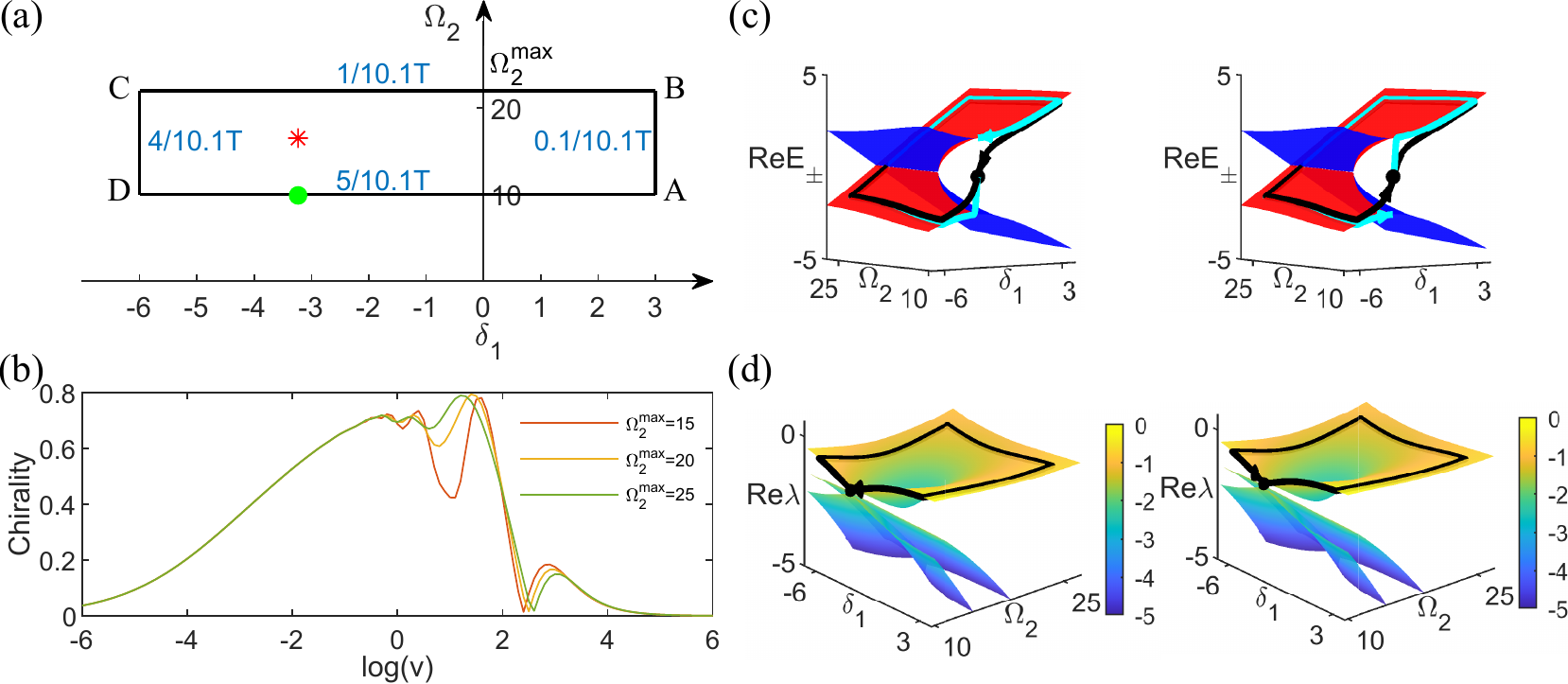}%
	\caption{(a) A more general encircling path, with no purely Hermitian segments. (b) Chirality as a function of $\log(v)$ for various $\Omega_2^{\text{max}}$. Other parameters are the same as those in Fig.~\ref{fig5}. (c)(d) Trajectories against the real eigenspectra of the non-Hermitian Hamiltonian $H_{\text{eff}}$ (c) and the Liouvillian $\mathcal{L}$ (d) for clockwise rotations (left panel) and counterclockwise rotations (right panel) when $\log(v)=-6$. Calculations of the trajectories are the same as those in Fig.~\ref{fig5}.
		\label{fig6}}
\end{figure*}

\section{Experimental relevance}

In Ref.~\cite{ca}, the collective chiral state transfer of an ultracold gas of fermions is observed, where the fermionic atom are subject to a synthetic spin-orbit coupling and laser-induced atom loss. For atoms that remain in the system, their dynamics is driven by a non-Hermitian effective Hamiltonian with an EP embedded in the parameter space. The laser-induced loss, however, is accompanied by spontaneous decay back into the system, similar to the $|3\rangle\rightarrow |2\rangle$ decay channel in Fig.~\ref{fig1}. Such a decay channel inevitably introduces decoherence to the system, which, according to our analysis above, should significantly impact the encircling dynamics.

To understand the experimental observation from the perspective of the master equation (\ref{eq1}), we replace $H_0$ in Eq.~(\ref{eq1}) as
\begin{align}
H_0=\begin{bmatrix}
	\begin{array}{cc}
		\frac{\delta_1}{2}-2q_x & -\Omega _1 \\
		-\Omega _1^* & -\frac{\delta_1}{2}+2q_x
	\end{array}
\end{bmatrix},
\end{align}
which corresponds to the Hermitian part of the Hamiltonian in Ref.~\cite{ca}, within the momentum sector $q_x$ (the $q_x^2$ terms are dropped since they are the same for the two spin components).
For our calculations, we take the recoil energy $E_r=2\pi \times 1.41$~kHz as the unit of energy, $k_r=\sqrt{2m E_r}$ ($m$ is the atomic mass) as the unit of momentum, $t_r=1/E_r$ as the unit of time. We focus on the momentum sector $q_x/k_r=-0.81$, as is the location of the Fermi surface under typical experimental conditions~\cite{ca}. Taking typical experimental parameters, we also fix $\gamma_0/E_r=110.57$, $\gamma_2/E_r=18.43$, $\Omega_1/E_r=-2.25$, and $\delta_2=0$ for our simulations below.

The experimental encircling path is reproduced in Fig.~\ref{fig5}(a) and the calculated chiralities under different encircling times are plotted in Fig.~\ref{fig5}(b). The initial parameters and state are, respectively, $\{\Omega_2/E_r=3$, $\delta_1/E_r=0\}$ (point $A$) and $\rho\left(0\right)=|2\rangle\langle2|$. For the counterclockwise encircling, we first ramp up $\Omega_2$ to $\Omega_2^{\text{max}}$ within the time $0.1/10.1 T$ ($A \rightarrow B$), and then decrease $\delta_1/E_r$ to $-6$ within $1/10.1 T$ ($B \rightarrow C$).
This is followed by decreasing $\Omega_2/Er$ to $0$ within $4/10.1 T$ ($C \rightarrow D$), and ramping up $\delta_1/Er$ to $3$ ($D \rightarrow A$).
The clockwise encircling ($A\rightarrow D \rightarrow C \rightarrow B \rightarrow A$) is
the exact reverse process of the above.
Since all the parameters are tuned in a linear fashion,  we parameterize the encircling time by the encircling velocity $v=2\pi/\left(T/t_r\right)$. The long-time limit is thus toward the left of the plot in Fig.~\ref{fig5}(b), when $v$ approaches zero.
While the chirality peaks at intermediate velocities (or intermediate encircling times), it approaches unity when $v$ approaches zero (or long encircling time).
In Ref.~\cite{ca}, $\log \left(v\right)\approx -2.66 $, where the chirality is already approaching unity, as shown in Fig.~\ref{fig5}(b). Under the same parameters, ideal EP encircling dynamics can be observed on the spectral landscape of the non-Hermitian Hamiltonian, as illustrated in Fig.~\ref{fig5}(c).
Curiously, this observation seems to contradict our conclusions in the previous section.

However, we notice that along the path $AD$ in Fig.~\ref{fig5}(a), $\Omega_2=0$ and the system is purely Hermitian. The Liouvillian gap closes along this sector. This renders our previous discussion irrelevant, as the open-system dynamics is typically non-adiabatic on the eigenspectral landscape of the Liouvillian. This is confirmed in Fig.~\ref{fig5}(d).
In the long-time limit, dynamics along $BC$ and $AD$ are both adiabatic, as the system follows the quas-steady state (eigenstate) of the Liouvillian (Hermitian Hamiltonian).
Along the segments $AB$ and $CD$ however, the dynamics can be non-adiabatic near the points $D$ and $A$, respectively, because of the closing of the Liouvillian gap.

Upon closer examination, the observed chirality originates from the distinct eigenstate overlaps between those of the Hermitian Hamiltonian and the Liouvillian near the gap-closing points $A$ and $D$. For the convenience of discussion, we denote the two eigenstates of the Hermitian Hamiltonian as $|\psi_\pm\rangle$, where $|\psi_+\rangle$ ($|\psi_-\rangle$) has a larger (smaller) eigenvalue.
The system is initialized in the state $|2\rangle$ at point $A$, which is close (on the spectral landscape) to $|\psi_-\rangle$. For a counterclockwise rotation ($A\rightarrow B \rightarrow C \rightarrow D \rightarrow A$) starting in the state $\rho\left(0\right)=|2\rangle\langle 2|$ at point $A$, the density matrix is in the quasi-steady state of the Liouvillian when approaching the gap-closing point $D$. Since the local quasi-steady state has a large overlap with $|\psi_-\rangle$ near point $D$, the system is projected onto a mixed state with considerable overlap with $|\psi_-\rangle$ upon the gap closing.
Following the adiabatic evolution along $DA$, the final state is then a mixed state that is close to $|\psi_-\rangle$ at point $A$, and is also close to the initial state.
By contrast, for a clockwise rotation ($A\rightarrow D \rightarrow C \rightarrow B \rightarrow A$), the quasi-steady state is projected onto a mixed state that is close to $|\psi_+\rangle$ near point $A$ along $BA$. The final state is therefore nearly orthogonal to the initial state.

We then adopt a different, and more general, encircling path, as shown in Fig.\ref{fig6}. As expected, the chirality drops to zero again in the long-time limit. This leaves an intermediate encircling time as the only window to observe the chiral state transfer in the presence of decoherence.

\section{Conclusion}
We study the impact of dephasing on the chiral state transfer near EPs in quantum open systems. We show that quantum jumps responsible for dephasing open up a Liouvillian gap, making the dynamics in the long-time limit necessarily adiabatic. Chiral state transfer thus typically emerges at intermediate times as an intrinsically non-adiabatic process. We discuss recent experiments in the light of our results. Specifically, in Ref.~\cite{sq1,sq2}, the observed chiral transfer in the superconducting qubits belong to the non-adiabatic category, which would disappear under a longer encircling time. On the other hand, the cold-atom experiment~\cite{ca} observes chiral transfer at long times thanks to the special path therein where the Liouvillian gap closes.
Throughout our work, we consider only the intrinsic dynamics, assuming a decoupling between the external (momentum) and internal degrees of freedom. For an atomic gas with lossy spin-orbit coupling as in Ref.~\cite{ca}, the recoil momentum accompanying the dephasing (quantum jump) process should also be considered for a better description of the system, particularly in regards to heating. We leave these considerations to future studies.
Our results provide a unified understanding for EP encirclings in quantum open systems, and are of direct relevance to on-going experimental efforts and quantum device design under realistic conditions.

\begin{acknowledgments}
We thank Gyu-Boong Jo for helpful communications. This work has been supported by the Natural Science Foundation of China (Grant No. 11974331) and the National Key R\&D Program (Grant No. 2017YFA0304100).
\end{acknowledgments}

\begin{widetext}

\appendix*

\section*{Appendix}

\section{Derivation of Equation~(\ref{eq1})}
Dynamics of system illustrated in Fig.~\ref{fig1} is governed by the Lindblad master equation
     \begin{align}
     \dot{\rho}=-i\left[H,\rho\right]+\sum_{i=0,1}\left[L_i \rho L_i^{\dagger}-\frac{1}{2}\left\{L_i^{\dagger}L_i,\rho\right\}\right],
     \end{align}
where $L_0=\sqrt{\gamma_0}\left|4\right\rangle\left\langle3\right|$, $L_1=\sqrt{\gamma_2}\left|2\right\rangle\left\langle3\right|$ . In the rotating frame, with rotating wave approximation the equations of matrix elements are
\begin{align}
    	\begin{cases}
    		\dot{\rho}_{11}&=i \Omega _1 \rho _{21}-i \rho _{12}\Omega _1^*\\
    		\dot{\rho}_{12}&=-i \left[\rho _{13} \Omega _2^*+\delta _1 \rho _{12}+\Omega _1 \left(\rho _{11}-\rho _{22}\right)\right]\\
    		\dot{\rho}_{13}&=-\rho _{13} \left(\gamma +i \delta _1+i \delta _2\right)-i \Omega _2 \rho _{12}+i \Omega _1 \rho _{23}\\
    		\dot{\rho}_{21}&=i \left[\rho _{31} \Omega _2+\delta _1 \rho _{21}+\Omega _1^*\left(\rho _{11}-\rho _{22}\right)\right]\\
    		\dot{\rho}_{22}&=\gamma _2 \rho _{33}+i \left[\rho _{12}\Omega _1^*-\rho _{23} \Omega _2^*-\Omega _1 \rho _{21}+\Omega _2 \rho _{32}\right]\\
    		\dot{\rho}_{23}&=i \left[i \rho _{23} \left(\gamma +i \delta _2\right)+\rho _{13} \Omega _1^*+\Omega _2 \left(\rho _{33}-\rho _{22}\right)\right]\\
    		\dot{\rho}_{31}&=-\rho _{31} \left(\gamma -i \delta _1-i \delta _2\right)+i\Omega _2^* \rho _{21}-i\Omega _1^* \rho _{32}\\
    		\dot{\rho}_{32}&=-i \left[-i \rho _{32} \left(\gamma -i \delta _2\right)+\rho _{31} \Omega _1+\Omega _2^* \left(\rho _{33}-\rho _{22}\right)\right]\\
    		\dot{\rho}_{33}&=i \left(2 i \gamma  \rho _{33}+\rho _{23} \Omega _2^*-\Omega _2 \rho _{32}\right)
    	\end{cases}
    \end{align}
with $\gamma=\left(\gamma_0+\gamma_2\right)/2$. The dynamics involving the state $\left|4\right\rangle$ is decoupled from that in the subspace of $\{|1\rangle,|2\rangle,|3\rangle\}$.

Assuming $\gamma\gg\left|\Omega_{1,2}\right|$, we set $\dot{\rho}_{13}=0$, $\dot{\rho}_{23}=0$, $\dot{\rho}_{33}=0$, $\dot{\rho}_{31}=0$, $\dot{\rho}_{32}=0$. We then have
      \begin{align}
     	\begin{cases}
     		\dot{\rho}_{11}&=-i \rho _{12}\Omega _1^*+i \Omega _1 \rho _{21} \\
     		\dot{\rho}_{12}&=-i \rho _{12} \left\{\delta _1-\frac{i \left| \Omega _2\right| {}^2 \left\{2 \gamma  \left(\gamma -i \delta _1-i \delta _2\right) \left(\left| \Omega _2\right| {}^2+\gamma ^2+\delta _2^2\right)+\left| \Omega _1\right| {}^2 \left[\left| \Omega _2\right| {}^2+2 \gamma  \left(\gamma +i \delta _2\right)\right]\right\}}{2 \gamma  \Delta^4}\right\}\\&+i \Omega _1 \rho _{22} \left\{1-\frac{\left| \Omega _2\right| {}^2 \left[\left| \Omega _1\right| {}^2+\left(\gamma -i \delta _2\right) \left(\gamma -i \delta _1-i \delta _2\right)\right]}{\Delta^4}\right\}+\frac{\Omega _1^2 \left| \Omega _2\right| {}^4}{2 \gamma  \Delta^4}\rho _{21}-i \Omega _1 \rho _{11}\\
     		\dot{\rho}_{21}&=i \rho _{21} \left\{\delta _1+\frac{i \left| \Omega _2\right| {}^2 \left\{2 \gamma  \left(\gamma +i \delta _1+i \delta _2\right) \left(\left| \Omega _2\right| {}^2+\gamma ^2+\delta _2^2\right)+\left| \Omega _1\right| {}^2 \left[\left| \Omega _2\right| {}^2+2 \gamma  \left(\gamma -i \delta _2\right)\right]\right\}}{2 \gamma  \Delta^4}\right\}\\&-i \Omega _1^* \rho _{22} \left\{1-\frac{\left| \Omega _2\right| {}^2 \left[\left| \Omega _1\right| {}^2+\left(\gamma +i \delta _2\right) \left(\gamma +i \delta _1+i \delta _2\right)\right]}{\Delta^4}\right\}+\frac{\left| \Omega _2\right| {}^4 \left(\Omega _1^*\right){}^2}{2 \gamma  \Delta^4}\rho _{12} +i \rho _{11} \Omega _1^*\\
     		\dot{\rho}_{22}&=-\frac{\gamma_0\rho _{22} \left| \Omega _2\right| {}^2 \left[\left| \Omega _1\right| {}^2+\gamma ^2+\left(\delta _1+\delta _2\right){}^2\right]}{\Delta^4}+i \Omega _1 \rho _{21} \left\{-1+\frac{\left(2 \gamma -\gamma _2\right) \left| \Omega _2\right| {}^2 \left[\left| \Omega _1\right| {}^2+\left(\gamma +i \delta _2\right) \left(\gamma +i \delta _1+i \delta _2\right)\right]}{2 \gamma  \Delta^4}\right\}\\&-i \rho _{12}\Omega _1^* \left\{-1+\frac{\left(2 \gamma -\gamma _2\right)\left| \Omega _2\right| {}^2 \left[\left| \Omega _1\right| {}^2+\left(\gamma -i \delta _2\right) \left(\gamma -i \delta _1-i \delta _2\right)\right]}{2 \gamma  \Delta^4}\right\}
     		\label{app1}
     	\end{cases}
      \end{align}
where
\begin{align}
      	\Delta ^4=\left| \Omega _1\right| {}^2 \left[\left| \Omega _2\right| {}^2+2 \gamma ^2-2 \delta _2 \left(\delta _1+\delta _2\right)\right]+\left[\gamma ^2+\left(\delta _1+\delta _2\right){}^2\right] \left(\left| \Omega _2\right| {}^2+\gamma ^2+\delta _2^2\right)+\left| \Omega _1\right| {}^4.
      \end{align}
Since $\Delta^4\sim \gamma^4$, we keep terms at least of the order of $\gamma^{-1}$ in Eq.~(\ref{app1}) to get
      \begin{align}
      	\dot{\rho}=-i
      	\begin{bmatrix}
      		\begin{array}{cc}
      			-\Omega _1 \rho _{21}+ \rho _{12}\Omega _1^* & \left(\delta_1-\frac{i\gamma ^3  \left| \Omega _2\right| {}^2}{\Delta^4}\right)\rho _{12}+\Omega _1 \left(\rho _{11}-\rho _{22}\right) \\
      			-\Omega _1^* \left(\rho _{11}-\rho _{22}\right)-\left(\delta_1+\frac{i\gamma ^3 \left| \Omega _2\right| {}^2}{\Delta^4}\right) \rho _{21} & -\frac{i\gamma _0 \gamma ^2 \rho _{22} \left| \Omega _2\right| {}^2}{\Delta^4}- \rho _{12} \Omega _1^*+ \Omega _1 \rho _{21} \\
      		\end{array}
      	\end{bmatrix}.
      \end{align}

This equation can be expressed as
     \begin{align}
     	\dot{\rho}=-i(H_{\text{eff}}\rho-\rho H_{\text{eff}}^{\dagger})+L_{\phi}\rho L_{\phi}^{\dagger}-\frac{1}{2}L_{\phi}^\dag L_{\phi}\rho-\frac{1}{2}\rho L_{\phi}^\dag L_{\phi},
     \end{align}
with $L_{\phi}=\sqrt{\gamma_{\phi}}\left|2\right\rangle\left\langle 2\right|$, $\gamma_{\phi}=\frac{\gamma_2\gamma ^2  \left| \Omega _2\right| {}^2}{\Delta ^4}$ and
     \begin{align}
     	H_{\rm{eff}}=
     	\begin{bmatrix}
     		\begin{array}{cc}
     			\frac{\delta_1}{2} & -\Omega _1 \\
     			-\Omega _1^* & -\frac{\delta_1}{2}-\frac{i\gamma_0\gamma^2  \left| \Omega _2\right| {}^2}{2\Delta ^4}
     		\end{array}
     	\end{bmatrix}=H_0-i\Gamma |2\rangle\langle 2|,
     \end{align}
 where $\Gamma=\frac{\gamma_0\gamma^2  \left| \Omega _2\right| {}^2}{2\Delta ^4}$.
This gives us Eq.~(\ref{eq1}) in the main text.
   \end{widetext}

\end{document}